\documentclass[twocolumn]{aastex631}
\usepackage{multirow}
\usepackage{makecell}
\usepackage{float}
\usepackage{needspace}
\usepackage{graphics}

\graphicspath{{./}{figures/}}

\begin{document}

\title{Discrepancy in Grain Size Estimation of H${}_{2}$O Ice in the Outer Solar System \protect\\ and the Interstellar Medium}


\correspondingauthor{A. Emran}
\email{alemran@uark.edu}

\author[0000-0002-4420-0595]{A. Emran}
\affiliation{Space and Planetary Sciences, University of Arkansas, Fayetteville, AR 72701, USA}

\author[0000-0002-1111-587X]{ V. F. Chevrier}
\affiliation{Space and Planetary Sciences, University of Arkansas, Fayetteville, AR 72701, USA}

\begin{abstract}
Widespread detection of amorphous and crystalline water (H${}_{2}$O) ice in the outer solar system bodies and the interstellar medium has been confirmed over the past decades. Radiative transfer models (RTMs) are used to estimate the grain sizes of H${}_{2}$O ice from near-infrared (NIR) wavelengths. Wide discrepancies in the estimation of H${}_{2}$O ice grain size on the Saturnian moons \citep{hansen2009}, as well as nitrogen (N${}_{2}$) and methane (CH${}_{4}$) ices on Kuiper belt objects have been reported owing to different scattering models used \citep{emran2022}. We assess the discrepancy in the grain size estimation of H${}_{2}$O ice at a temperature of 15, 40, 60, and 80 K (amorphous) and 20, 40, 60, and 80 K (crystalline) -- relevant to the outer solar system and beyond. We compare the single scattering albedos of H${}_{2}$O ice phases using the Mie theory \citep{mie1908} and Hapke approximation models \citep{hapke1993} from the optical constant at NIR wavelengths (1 -- 5 µm). This study reveals that the Hapke approximation models -- Hapke slab and internal scattering model (ISM) -- predict grain size of the crystalline phase, overall, much better compared to the amorphous phase at temperatures of 15 -- 80 K. However, the Hapke slab model estimates much approximate grain sizes, in general, to that of the Mie model's prediction while ISM exhibits a higher uncertainty. We recommend using the Mie model for unknown spectra of outer solar system bodies and beyond in estimating H${}_{2}$O ice grain sizes. While choosing the approximation model for employing RTMs, we recommend using a Hapke slab approximation model over the internal scattering model.
\end{abstract}



\section{Introduction} \label{sec:intro}

Detection of water (H${}_{2}$O) ice on outer solar system bodies (giant planets' rings and their icy moons and dwarf planets), cometary bodies (originates from the Kuiper belt and Oort cloud), and interstellar medium are well established (see \citealt{kofman2019} for an exhaustive list of references). Solid H${}_{2}$O ice in these environments may appear as either crystalline or amorphous or both phases based on temperature, radiation histories, and their formation temperature and pressure conditions \citep{mastrapa2008}. Though amorphous water ice (alternatively microscopic solid H${}_{2}$O ice) is considered the most abundant form of H${}_{2}$O in the universe  \citep{baragiola2003}, the crystalline phase has also been spotted in the many outer solar system bodies.

Phases of amorphous and crystalline H${}_{2}$O ice were investigated on the Galilean moons of Jupiter using the data from NIMS instrument onboard Galileo spacecraft \citep{hansen2004}. Neptune's moon Triton, a trans-Neptunian object (TNO), is also assumed to host amorphous and crystalline water ice \citep{cruikshank2000}. Although the presence of crystalline H${}_{2}$O ice has undoubtedly been reported, in a small fraction, on Pluto at the Kuiper belt \citep{cook2019}, its largest satellite Charon hosts both amorphous and crystalline H${}_{2}$O phases \citep{dalle2018}. Molecules of amorphous H${}_{2}$O ice have been detected on the icy dust grains in dense interstellar clouds \citep[e.g.,][]{herbst2001}. Due to the formation of cometary icy grains in very low temperatures, the water ice within cometesimals is thought to be in an amorphous phase \citep{raponi2016}. However, the phase transition of H${}_{2}$O ice from amorphous to crystalline on the cometary surface happens when comets pass a closer distance to the Sun which leads to an activation temperature of 120 K for phase transition \citep[e.g.,][]{raponi2016}.

Observations from different spectral wavelengths are used to study the outer reach of the solar system. Among the wavelengths, the near-infrared (NIR) has the most diagnostic bands and therefore have extensively been used for characterizing ices and volatiles in outer solar system bodies \citep{barucci2020}. While the observations of ices in the interstellar medium are mostly accomplished from the spectral absorption signatures of infrared lights from reddened stars \citep{kofman2019}, the H${}_{2}$O ice phases in the interstellar medium can remotely be determined by observing the 3 µm band \citep{mastrapa2009l}. The interaction between infrared photons and H${}_{2}$O ices and their resultant absorption bands and positions are dependent on its crystalline vs amorphous phases and the temperature of the ice \citep{mastrapa2009l}. Thus, the shapes and position of the infrared absorption bands and their variations in response to temperature are considered as the diagnostic for astronomical ices \citep[e.g.,][]{schmitt1998}. At NIR wavelengths, amorphous water ice exhibits different spectral characteristics above and below the deposition temperature at 70 K such that the amorphous ice has indistinguishable shapes in spectral bands below that temperature \citep{mastrapa2008}.

Radiative transfer models (RTMs) have been used for characterizing the composition of the outer solar system bodies such as Kuiper belt objects (KBOs) and TNOs \citep[e.g.,][]{grundy1991, dumas2007, merlin2010, tegler2010} and molecular clouds and accretion disk \citep[e.g.,][]{pollack1994}. Wide discrepancies in the estimation of grain size have been reported for H${}_{2}$O ice on Saturnian moons such as Enceladus \citep{hansen2009} and nitrogen (N${}_{2}$) and methane (CH${}_{4}$) ices on TNOs and KBOs using different scattering models \citep{emran2022}. This uncertainty in grain sizes of ices is due to the use of different single scattering albedo (\textit{w}) calculations rather than a choice of using bidirectional scattering models \citep{hansen2009}. Accordingly, \cite{hansen2009} compares the single scattering albedos using the optical constant data of crystalline H${}_{2}$O ice at 110 K -- relevant to Saturnian moons and other icy bodies. However, temperatures at the far reach of the solar system at the Kuiper belt, Oort cloud, and the interstellar medium are much lower. For instance, Pluto has a mean temperature of $\mathrm{\sim}$ 40 K \citep[e.g.,][]{earle2017} while the temperatures of molecular clouds in the interstellar environment of our Milky Way galaxy are assumed to be 10 - 20 K \citep[e.g.,][]{ferriere2001}.

Moreover, while employing RTMs in the outer solar system bodies, there has been extensive use of optical constants of crystalline and/or amorphous phases of H${}_{2}$O ice and a variety of scattering models. For instance, \cite{protopapa2017} used the optical constant of crystalline H${}_{2}$O ice and employed the equivalent slab model of \cite{hapke1993} for mapping the spatial distribution and grain size of water ice (along with other ices) on Pluto. For the same planetary body, \cite{cook2019} used the optical constant of both amorphous and crystalline ices while their implemented scattering model was following the formulations of \cite{roush1994} and \cite{cruikshank1998} with some modifications. Thus, the uncertainty of the grain size estimation of amorphous and crystalline water ice at temperatures relevant to the outer solar system and the interstellar medium (i.e., 15 K -- 80 K) using different scattering models is warranted.

We assess the uncertainty in predicting the H${}_{2}$O ice grain sizes at a temperature of 15, 40, 60, and 80 K (amorphous) and 20, 40, 60, and 80 K (crystalline) relevant to the outer solar system bodies including the Kuiper belt, Oort cloud, and the interstellar medium. Single scattering albedos as being the primary cause of the varied results in the grain size estimation of ices \citep{hansen2009}, we compare the relative grain sizes of H${}_{2}$O ice predicted by different single scattering models. The theory of \cite{mie1908} and approximation models of \cite{hapke1993} are commonly used methods for calculating the \textit{w} of a material from its optical constants. Accordingly, we calculate the \textit{w} of water (amorphous and crystalline) ices by implementing the Mie theory and widely used Hapke approximation models from the optical constants over the wavelengths between 1 - 5 µm.

\section{Methods} \label{sec:methods}

\subsection{Optical constants}
We use the NIR optical constants of H${}_{2}$O ice at a temperature of 15 -- 80 K (amorphous phase) and 20 -- 80 K (crystalline phase). The optical constants of amorphous and crystalline ice at NIR bands between 1.1 -- 2.6 µm were collected from \cite{mastrapa2008} and NIR to mid-infrared bands between 2.5 -- 22 µm from \cite{mastrapa2009l}. A compiled optical constant data for both ice phases over the entire NIR wavelengths (1 -- 5 µm) can be found in \cite{mastrapa2009l}. Note that the crystalline phase used in this study is assumed to be the cubic ice form -- a meta-stable H${}_{2}$O ice phase \citep{mastrapa2009l}. The cubic and hexagonal crystalline ices exhibit nearly identical spectra at infrared bands \citep{bertie1967}, and in this study, we do not distinguish our discussion on cubic vs hexagonal ice form separately rather we treat the crystalline phases as a unified entity. We use the cubic ice as the representative of the crystalline ice, and hereafter, we denote the crystalline ice as Ic and amorphous ice as Ia.

We use the optical constant data from \cite{mastrapa2008, mastrapa2009l} to maintain consistency as prepared by the same instrumental setup and author(s), though there have been other sources for optical constants of water ice elsewhere too. For instance, \cite{grundy&schmitt1998} prepared the optical constants for monocrystalline (hexagonal) H${}_{2}$O ice samples between the wavelengths of 0.97 -- 2.73 µm at temperatures of 20 to 270 K. On the other hand, \cite{hudgins1993} prepared the optical constants of Ia between the wavelengths of 2.5 -- 20 µm at temperatures of 10 -- 120 K.

\subsection{ Single scattering albedo}

Photon interaction with a material grain happens at the atomic-molecular level through the process of selective absorption, emission, and scattering \citep[e.g.,][]{shepard2017}. The sum of photons scattered from or absorbed by a particle to the number of photons, at a given wavelength, by its physical cross-section is called extinction (or attenuation) efficiency, \textit{Q${}_{ext\ }$= Q${}_{sc}$ + Q${}_{ab}$}${}_{.\ }$The \textit{w} is the ratio of scattering efficiency and extinction efficiency, \textit{w} = \textit{Q${}_{sc\ }$/ Q${}_{ext}$}.

For planetary regolith\textit{, w} is a representative of characteristic surface properties including grain size, its optical properties, internal structure, and partly shapes \citep{hapke1981}. Consequently, the grain size estimation of the regolith primarily depends on the characteristic value of \textit{w} by the surface particles \citep{hansen2009}. Material's \textit{w} is a function of optical properties (indices of refraction) of its single grain \citep[e.g.,][]{hapke1993, mustard&glotach2019}. Thus, \textit{w} can be calculated if optical constants -- real (\textit{n}) and imaginary (\textit{k})\textit{ }parts -- of regolith medium are known. Larger particles with moderate to larger \textit{k} values result in higher absorption of incident light and therefore, are associated with a lower \textit{w} value \citep[e.g.,][]{shepard2007}.

Mie scattering theory can accurately estimate the \textit{w} for a particle with a spherical shape \citep{mie1908, wiscombe1980}. On the other hand, \cite{hapke1993} formulates a variety of approximation models for the estimation of \textit{w} from material optical constants. We employed two of these approximation models -- the ``\textit{espat}'' slab (hereafter the Hapke slab or simply slab model; \citealt{hapke1981, hapke1993}) and a version of the internal scattering model (ISM; \citealt{hapke1993}) -- which have widely been used in the RTMs of the outer solar system bodies. Note that the version of the scattering model by \cite{roush1994} uses, indeed, a formulation of the ISM in the original paper by \cite{hapke1981} and devised the versions of surface scattering functions \citep{hansen2009}.

\subsection{Mie calculation}

The scattering properties of a particle with the simplest three-dimensional geometrics (spheres) can accurately be calculated using the Mie theory \citep{mie1908} by employing Maxwell's equations. The formulation of Mie theory accounts for optical constants (\textit{n} + \textit{ik}) of the particle and the fraction of particle's size to wavelengths under investigation \citep{hapke1993}. However, the treatment of diffraction effects in Mie's \textit{w} is required for a scattering model with highly asymmetric phase functions \citep{hansen2009}. Thus, in this study we use \textit{$\delta$}-Eddington corrected \citep{joseph1976} Mie single scattering albedo,\textit{ w'}, calculated as \citep{wiscombe&warren1980}:

\begin{equation} \label{Eq.1} 
w'=\frac{\left(1-{\xi}^2\right)\ w}{1-{\xi}^{2\ }w}
\end{equation}

where \textit{$\xi$} is the asymmetry factor calculated by Mie theory. Following the approach of \cite{emran2022}, the Mie \textit{w} was calculated by utilizing the method of \cite{wiscombe1979} using the {\tt\string miepython} routine (an open-source Python module). The \textit{$\delta$}-Eddington corrected Mie \textit{w'} was then adjusted from Mie's \textit{w} following Eq. \ref{Eq.1}.

\subsection{Hapke approximation models}

Using the ``\textit{espat}'' function or the slab model of \cite{hapke1981, hapke1993}, the approximate \textit{w} of an equant particle can be simplified as:

\begin{equation} \label{Eq.2} 
w=\frac{1}{\alpha D_e+1}
\end{equation} 

where \textit{$\alpha$ }is the absorption coefficient and \textit{D${}_{e}$} is the ``effective particle size''. The absorption coefficient is given by:

\begin{equation} \label{Eg.3} 
\alpha =\frac{4\pi k}{\lambda }
\end{equation}

where \textit{k} is the imaginary part of the optical constant and \textit{$\lambda$} is the wavelength under consideration. The Hapke slab model should be applied for materials with \textit{k $<<$}1 and \textit{D${}_{e}$} can be approximated as \citep{hapke1993}:

\begin{equation} \label{Eq.4} 
D_e=\mathrm{\textrm{\^{g}}}\frac{1-S_e}{1-S_i} \textit{D}  
\end{equation} 

where \textit{D} is the particle diameter, \textit{\^{g}} is a constant (scaling parameter of diameter) which is roughly $=$  1, and \textit{S${}_{e}$} and \textit{S${}_{i}$} are the average Fresnel reflection coefficient for externally and internally incident light, respectively. The derivations of the approximate equations for the calculation of \textit{S${}_{e}$} and \textit{S${}_{i}$} are given in \cite{hapke1993} where \textit{k} is set to be zero, therefore is not considered, for the slab model. These approximate equations (\textit{S${}_{e}$} and \textit{S${}_{i}$}) can also be found elsewhere in the literature that uses the Hapke models including the study of \cite{emran2022}. The calculation of \textit{w} using the ISM can be approximated as \citep{hapke1993}:

\begin{equation} \label{Eq.5} 
w=S_e+\left(1-S_e\right)\frac{\left(1-S_i\right)\textrm{$\Theta$}}{1-S_i\textrm{$\Theta$}}\  
\end{equation} 

where particles’ internal-transmission function, \textit{$\Theta$} is given by \citep{hapke1993}:

\begin{equation} \label{Eq.6} 
\textrm{$\Theta$}=\frac{r+\mathrm{exp}\mathrm{}(-\sqrt{}(\alpha \left(\alpha +s\right))\langle D\rangle }{1-r\ \mathrm{exp}\mathrm{}(-\sqrt{}\left(\alpha \left(\alpha +s\right)\right)\langle D\rangle } 
\end{equation}

where \textit{s} is the near-surface internal scattering coefficient. The mean free path of photon \textit{$\langle $D$\rangle $} as a function of \textit{n} for perfectly spherical particle can be written as  \citep{hapke1993}:

\begin{equation} \label{Eq.7} 
\langle D\rangle \ =\frac{2}{3}\left(n^2-\frac{1}{n}{\left(n^2-1\right)}^{\frac{3}{2}}\right)D 
\end{equation} 

The internal hemispherical (diffused) reflectance, \textit{r} can be given as:

\begin{equation} \label{Eq.8} 
r=\frac{1-\ \sqrt{}\alpha \left(\alpha +s\right)}{1+\ \sqrt{}\alpha \left(\alpha +s\right)} 
\end{equation}

If \textit{s} is set to zero \citep{lucey1998}, then \textit{r} = 0 and \textit{$\mathit{\Theta}$} = \textit{e${}^{-\ \alpha }$ ${}^{\langle }$${}^{D}$${}^{\rangle }$} \citep[e.g.,][]{lawrence2007, hansen2009, li2011}. The list of notations used in this study are given in Appendix (\ref{Notation}).

\section{Results}

\subsection{Calculated single scattering albedo}

\begin{figure*}[ht!]
\plotone{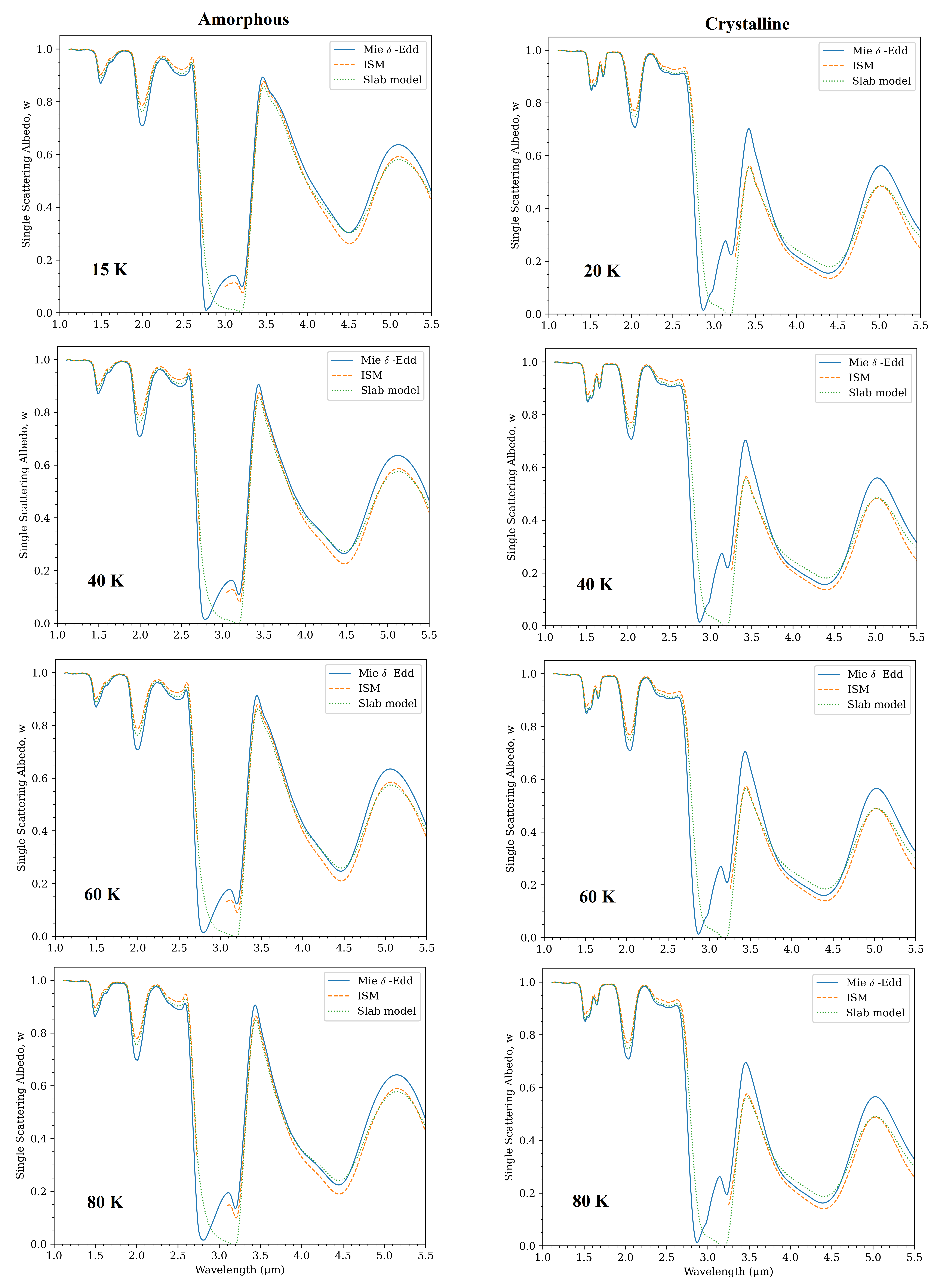}
\caption{Calculated single scattering albedo plots for 10 µm radii particles of amorphous (left column) and crystalline (right column) H${}_{2}$O ice at the NIR wavelengths. The subplots in the rows are the measurements at different temperatures where each dashed solid line is \textit{$\delta$}-Eddington corrected Mie, the dotted green is Hapke slab, and dashed yellow is ISM spectra.
\label{Fig1}}
\end{figure*}

We compare the \textit{w} spectra of 10 µm radii Ia and Ic particles to assess the difference in calculated albedo spectra at that specific diameter calculated using the Mie and Hapke approximation models over the NIR bands (Fig. \ref{Fig1}). Following the approach of \cite{emran2022}, we apply the \textit{Savitzky-Golay} filter \citep{savitzky1964} to smoothen the small spikes in calculated \textit{w} curves over the wavelengths. We implement the filter algorithm with a third order polynomial fit. Note that all plots used in this study were smoothened using the \textit{Savitzky-Golay} filter, if not mentioned otherwise.

The \textit{w} plots from the Hapke approximation models follow the spectra generated by the Mie model with varying degrees of closeness at different NIR wavelengths. In the case of Ia at every investigated temperature (the left column in Fig. \ref{Fig1}), the Hapke slab models follow much closer, overall, to the Mie model spectra compared to the ISM model spectra, except around the broader 3.0 µm region. At shorter wavelengths up to 2.5 µm, the slab model produces slightly higher \textit{w} values than the Mie model, notably around weak overtone 1.5 and 2.0 µm bands at every investigated temperature. This represents a prediction of slightly smaller H${}_{2}$O grains by the slab model than the Mie model \citep{hansen2009} over the shorter wavelengths at weak overtone 1.5 and 2.0 µm bands. Neither of the approximation model spectra does mimic the Mie spectra at 3 µm wavelengths, except the ISM spectra do partially follow the shape of the Mie spectra on the bands past 3 µm. The inability of the slab model in predicting the 3 µm absorption band is likely due to having higher \textit{k} values at the wavelengths.

For Ia at the broader 4.5 µm absorption band, the slab model either exactly follows the Mie spectra (at 15K) or renders slightly higher \textit{w} values than the Mie spectra (at 40, 60, and 80K). This represents that the slab model at these wavelengths either exactly predicts the same grain sizes or produces slightly smaller grain sizes than Mie's prediction, respectively. In contrast, at every temperature, the ISM renders much lower \textit{w} than the Mie spectra at 4.5 µm band -- indicating larger grain size prediction by the ISM. At longer wavelengths around 3.5 and 5 µm, both approximation models show smaller\textit{ w} values than the Mie model -- meaning a larger grain size prediction by the approximation models at these bands. However, at the latter band, the grain size prediction is larger than the former band.

In the instance of Ic at every temperature (the right column in Fig. \ref{Fig1}), the slab model accurately mimics the Mie model at the weak overtone 1.5 µm while it shows slightly higher \textit{w} values at the weak overtone 2.0 µm. This represents an accurate grain size prediction at 1.5 µm and a slightly smaller grain size prediction at 2.0 µm by the Hapke slab model. The ISM, on the other hand, predicts much smaller grain sizes at 1.5 and 2.0 µm. The reflection peak at 3.1 µm does not represent by either of the approximation models due to having a higher \textit{k} value at this band \citep{hansen2009}. At the 4.4 µm absorption band, the slab model predicts smaller grain sizes while the ISM predicts larger grain sizes compared to Mie's prediction. However, both the approximation models follow the same path and show much smaller \textit{w} values at 3.5 and 5 µm bands -- indicating a much larger grain size prediction by the approximation models. Compared to Ia, the spectra of Hapke approximation models in Ic show a much larger difference than Mie's value at 3.5 µm.

\Needspace{5\baselineskip}
\Needspace{5\baselineskip}
\subsection{Relative grain size estimation}

We assess the uncertainty in the predictions of grain sizes by the Hapke approximation models for the ice phases at the investigated temperatures. A finely spaced range of grain sizes at radii of 1, 10, 100, and 1000 µm was used to show the relative discrepancies of grain size prediction at NIR bands. We choose this finely spaced range because it covers the grain sizes, from microns (µm) to millimeters (mm), found in the outer solar system bodies. For instance, fine-grain H${}_{2}$O ice with a diameter of $\mathrm{<}$ 1 µm has been predicted on Pluto \citep{cook2019}. The presence of much larger ices grain sizes is also believed to be present in outer solar system bodies such as icy moons and rings of Saturn \citep[e.g.,][]{filacchione2012} and Pluto \citep[e.g.,][]{cook2019}.

Accordingly, we first compute the \textit{$\delta$}-Eddington corrected Mie \textit{w' }at these fixed grain radii over the NIR bands. Once the Mie \textit{w'} is \textit{ }calculated, we estimate the corresponding grain sizes from each of the Hapke approximation models by fitting the Mie spectra over the wavelengths. To this end, we solve the inverse of ISM and Hapke slab model for \textit{D} to fit the corresponding Mie \textit{w' }values. We use Powell's hybrid method \citep{powell1970, chen1981} to solve the non-linear function of Hapke slab and ISM (Eq. \ref{Eq.2} and \ref{Eq.5}) in finding the solution of \textit{D}. The relative grain size prediction was then determined by normalizing the predicted grain sizes by the approximation models to the grain sizes of Mie model.

\begin{figure*}[ht!]
\plotone{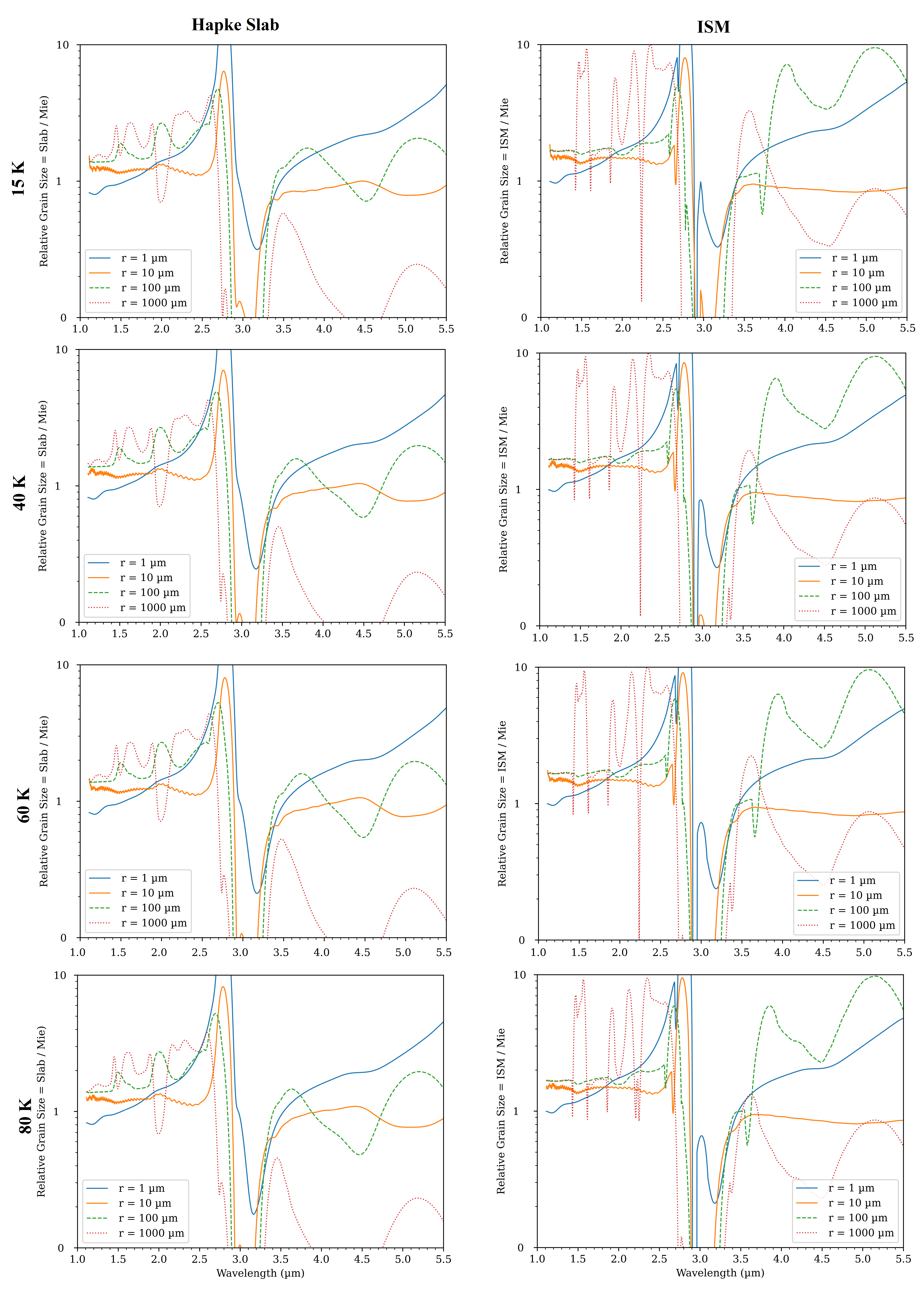}
\caption{Predicted relative gain sizes (normalized) using the Hapke slab (left column) and ISM (right column) calculated at particle radii of 1, 10, 100, and 1000 µm for amorphous H${}_{2}$O ice. The subplots in the rows are the measurements at different temperatures.
\label{Fig2}}
\end{figure*}

\begin{figure*}[ht!]
\plotone{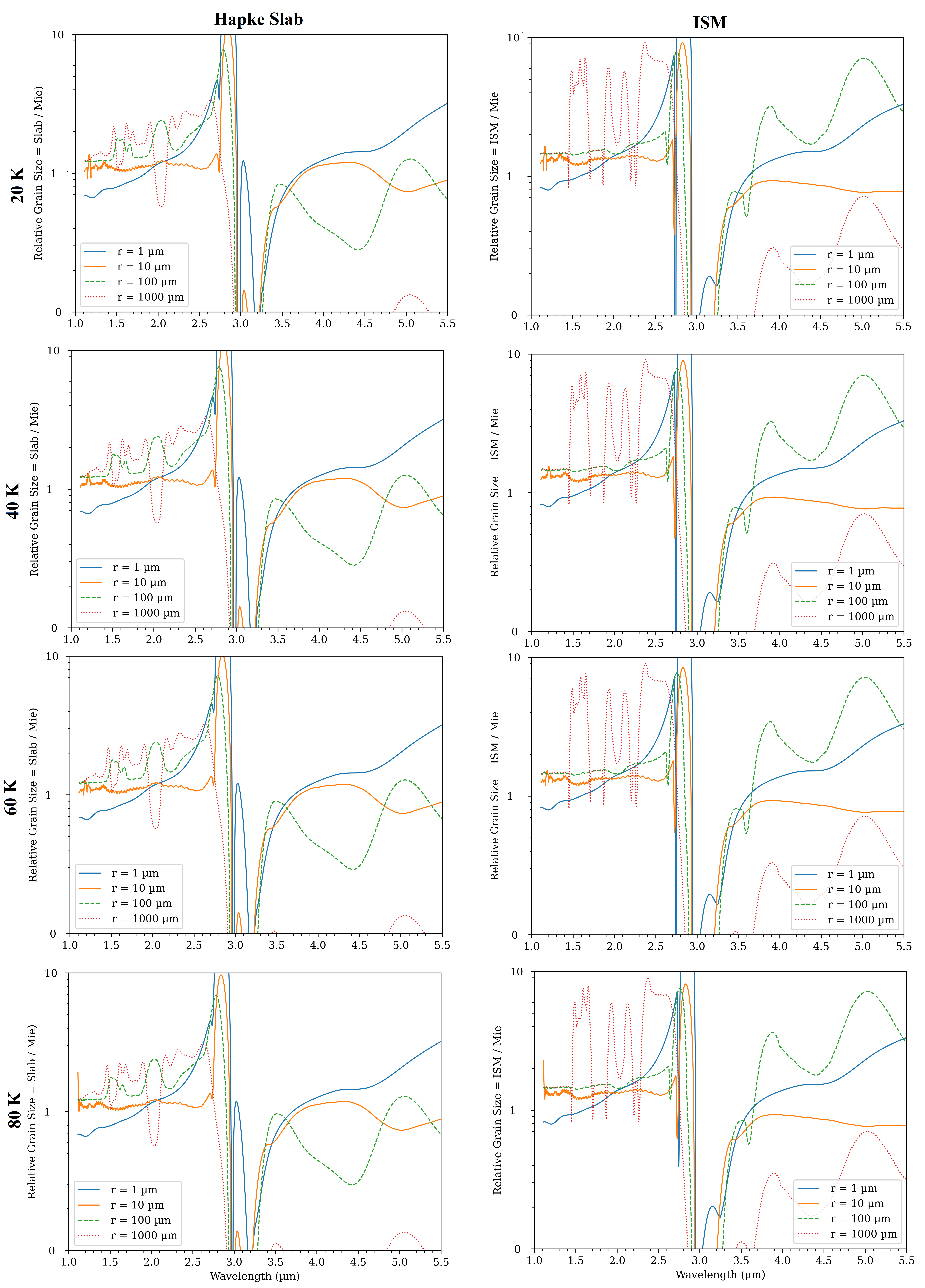}
\caption{Predicted relative gain sizes (normalized) using the Hapke slab (left column) and ISM (right column) calculated at particle radii of 1, 10, 100, and 1000 µm for crystalline H${}_{2}$O ice. The subplots in the rows are the measurements at different temperatures.
\label{Fig3}}
\end{figure*}

We estimate the statistics, medians and their 16\% quantile as a lower 1$\sigma$ error bar and 84\% quantile as an upper 1$\sigma$ error bar, of relative discrepancies in predicted grain sizes using the approximation models at different particle radii for Ia (Table \ref{tab:1}a) and Ic (Table \ref{tab:1}b) at the investigated temperatures. As additional supporting information, we also list the mean $\mathrm{\pm}$ 1$\sigma$ standard deviation relative discrepancies in predicted grain sizes in the Appendix (Table \ref{tab:2}). The statistics were derived from the normalized values of the estimated grain sizes predicted by the slab model and ISM to the Mie model. A value of 1.0 represents exact prediction while the higher or lower values imply an over or underestimation, respectively, of the grain sizes by the respective approximation models. The plots for predicted relative gain sizes using the Hapke slab and ISM for Ia and Ic are given in Fig. \ref{Fig2} and Fig. \ref{Fig3}, respectively.

\begin{table*}
\begin{center}

\large

\caption{The statistics - medians and their 16\% quantile as a lower 1$\sigma$ error bar and 84\% quantile as an upper 1$\sigma$ error bar - of relative discrepancies in predicted grain sizes using the Hapke slab and ISM calculated at particle radii of 1, 10, 100, and 1000 µm for amorphous (a) and crystalline (b) H${}_{2}$O ice at different temperatures. \label{tab:1}}

\begin{tabular}{lcccccccc}


\multicolumn{9}{l}{a.}\\
\hline

\multirow{3}{*}{\rotatebox{90}{
\makecell{Grain radii \hspace{.05in} \\ (µm)}}} & \multicolumn{8}{c}{\textbf{Amorphous H${}_{2}$O ice${}^{*}$}}  \\ [6pt]
\cline{2-9} 
 & \multicolumn{2}{c}{15 K} & \multicolumn{2}{c}{40 K} & \multicolumn{2}{c}{60 K} & \multicolumn{2}{c}{80 K} \\ [6pt]
\cline{2-9}
 & Slab & ISM & Slab & ISM & Slab & ISM & Slab & ISM \\  [6pt] \hline 
1 & ${\mathrm{1.07}}^{+1.04}_{-0.25}$ & ${\mathrm{1.28}}^{+1.08}_{-0.29}$  & ${\mathrm{1.07}}^{+0.97}_{-0.25}$  & ${\mathrm{1.28}}^{+1.02}_{-0.29}$  & ${\mathrm{1.07}}^{+0.96}_{-0.25}$  & ${\mathrm{1.29}}^{+1.04}_{-0.30}$  & ${\mathrm{1.08}}^{+0.92}_{-0.26}$  & ${\mathrm{1.29}}^{+1.03}_{-0.30}$ \\ \hline 
10 & ${\mathrm{1.19}}^{+0.06}_{-0.34}$ & ${\mathrm{1.44}}^{+0.05}_{-0.56}$  & ${\mathrm{1.20}}^{+0.07}_{-0.29}$  & ${\mathrm{1.45}}^{+0.05}_{-0.57}$  & ${\mathrm{1.21}}^{+0.06}_{-0.31}$  & ${\mathrm{1.46}}^{+0.05}_{-0.58}$  & ${\mathrm{1.21}}^{+0.06}_{-0.26}$  & ${\mathrm{1.45}}^{+0.05}_{-0.58}$  \\ \hline 
100 & ${\mathrm{1.46}}^{+0.58}_{-0.12}$  & ${\mathrm{1.66}}^{+0.35}_{-0.10}$  & ${\mathrm{1.46}}^{+0.58}_{-0.26}$  & ${\mathrm{1.66}}^{+0.48}_{-0.09}$  & ${\mathrm{1.47}}^{+0.62}_{-0.26}$  & ${\mathrm{1.67}}^{+0.45}_{-0.09}$  & ${\mathrm{1.47}}^{+0.60}_{-0.35}$  & ${\mathrm{1.67}}^{+0.51}_{-0.09}$  \\ \hline 
1000 & ${\mathrm{1.54}}^{+1.08}_{-1.37}$  & ${\mathrm{1.67}}^{+4.34}_{-1.08}$  & ${\mathrm{1.54}}^{+1.09}_{-1.38}$  & ${\mathrm{1.67}}^{+4.34}_{-1.19}$  & ${\mathrm{1.55}}^{+1.11}_{-1.37}$ & ${\mathrm{1.68}}^{+4.35}_{-1.23}$ & ${\mathrm{1.55}}^{+1.10}_{-1.40}$  & ${\mathrm{1.68}}^{+3.87}_{-1.26}$ \\ \hline 
& & & & & & & & \\

\multicolumn{9}{l}{ b. }\\
\cline{1-9}

\multirow{3}{*}{\rotatebox{90}{
\makecell{Grain radii \hspace{.05in} \\ (µm)}}} & \multicolumn{8}{c}{\textbf{Crystalline H${}_{2}$O ice${}^{*}$}}  \\ [6pt]
\cline{2-9} 
 & \multicolumn{2}{c}{20 K} & \multicolumn{2}{c}{40 K} & \multicolumn{2}{c}{60 K} & \multicolumn{2}{c}{80 K} \\ [6pt]
\cline{2-9}
 & Slab & ISM & Slab & ISM & Slab & ISM & Slab & ISM \\  [6pt] \hline 
1 & ${\mathrm{0.88}}^{+0.67}_{-0.19}$  & ${\mathrm{1.03}}^{+0.76}_{-0.22}$ & ${\mathrm{0.87}}^{+0.67}_{-0.19}$  & ${\mathrm{1.03}}^{+0.76}_{-0.22}$  & ${\mathrm{0.88}}^{+0.67}_{-0.19}$  & ${\mathrm{1.03}}^{+0.75}_{-0.22}$  & ${\mathrm{0.87}}^{+0.67}_{-0.19}$  & ${\mathrm{1.03}}^{+0.75}_{-0.22}$ \\ \hline 
10 & ${\mathrm{1.10}}^{+0.08}_{-0.07}$  & ${\mathrm{1.30}}^{+0.07}_{-0.44}$ & ${\mathrm{1.09}}^{+0.07}_{-0.06}$ & ${\mathrm{1.30}}^{+0.07}_{-0.44}$  & ${\mathrm{1.09}}^{+0.07}_{-0.07}$  & ${\mathrm{1.30}}^{+0.07}_{-0.44}$  & ${\mathrm{1.09}}^{+0.07}_{-0.07}$ & ${\mathrm{1.30}}^{+0.07}_{-0.44}$ \\ \hline 
100 & ${\mathrm{1.24}}^{+0.65}_{-0.54}$  & ${\mathrm{1.46}}^{+0.40}_{-0.03}$  & ${\mathrm{1.24}}^{+0.65}_{-0.54}$  & ${\mathrm{1.46}}^{+0.40}_{-0.03}$  & ${\mathrm{1.25}}^{+0.64}_{-0.51}$  & ${\mathrm{1.46}}^{+0.41}_{-0.03}$  & ${\mathrm{1.24}}^{+0.64}_{-0.51}$ & ${\mathrm{1.46}}^{+0.41}_{-0.03}$  \\ \hline 
1000 & ${\mathrm{1.34}}^{+0.80}_{-1.27}$ & ${\mathrm{1.47}}^{+3.57}_{-1.25}$  & ${\mathrm{1.34}}^{+0.81}_{-1.27}$  & ${\mathrm{1.67}}^{+3.64}_{-1.25}$ & ${\mathrm{1.34}}^{+0.81}_{-1.26}$  & ${\mathrm{1.47}}^{+3.78}_{-1.24}$  & ${\mathrm{1.34}}^{+0.81}_{-1.26}$ & ${\mathrm{1.47}}^{+3.78}_{-1.24}$ \\ \hline

\end{tabular}


\end{center}

\tablecomments{The statistics were computed by normalizing the estimated grain sizes from the Hapke slab and ISM to the Mie grain size. The number values are the medians and their 16\% quantile as a lower 1$\sigma$ error bar and 84\% quantile as an upper 1$\sigma$ error bar -- representing how many times of predicted grain sizes by slab and ISM models to the Mie grain size H${}_{2}$O ice phases at different temperatures}

\end{table*}

The predicted grain sizes by both approximation models are much closer, overall, to Mie's prediction for the Ic than the Ia. This is supported by the better-predicted medians values (much closer to 1) in the relative grain size estimation by the Ic compared to the Ia (Table \ref{tab:1}) -- also visually evident in Fig. \ref{Fig2} and \ref{Fig3}. For instance, between the ice phases at the corresponding grain radii (e.g., 10 µm) and temperatures (e.g., 40 K), the median values are much better by both approximation models for Ic compared to the Ia. An exception in this comparison is that for 1 µm grain the slab model’s prediction is slightly closer to 1 for Ia than Ic; however, their 16 and 84\% quantile error bars are higher. The mean values (Table \ref{tab:2} in Appendix) between the ice phases are also consistent -- although not completely at every temperature and grain size -- with the median values, except for smaller grains ($\mathrm{\le}$ 10 µm) at lower temperatures ($\mathrm{\le}$ 40 K). Therefore, it can be concluded that the Hapke approximation models predict much better, in general, grain sizes for the Ic than the Ia phase.

Between the approximation models, the slab model shows better approximations (much closer to 1) than the ISM for both ice phases -- evident from the median (except for Ic at 1 µm) and mean $\mathrm{\pm}$ 1$\sigma$ values at every corresponding temperature and grain size. Even though the slab model-predicted median values for 1 µm particles are slightly farthest for Ic, their 16 and 84\% quantile error bars indicate a lower discrepancy than the ISM. The plots for the relative discrepancies in grain size of Ia (Fig. \ref{Fig2}) and Ic (Fig. \ref{Fig3}) also support that the slab model predicts a much better approximate result. In contrast, the discrepancies in grain size estimation by ISM are much higher -- consistent with the result of \cite{hansen2009}. However, around the absorption peak of the 3.1 µm band, neither of the approximation models predicts good results.

The 1 µm curve at every temperature and ice phase appears a continuous rise at shorter wavelengths up to $\mathrm{\sim}$2.8 µm. This continuous rise at shorter bands is largely due to the Rayleigh effect on \textit{w} from the Mie model that happens at wavelengths close to or smaller than the grain size \citep{hansen2009}. The Hapke approximation models do not account for the Rayleigh effect of scattering models. Thus, even if the median values of the 1 µm curve show a good prediction (Table \ref{tab:1}), the mean and 1$\sigma$ standard deviation in predicated grain sizes (Table \ref{tab:2} in Appendix) show one of the highest discrepancies in grains size estimation by the approximation models. The 10 µm curve for each ice phase and temperature illustrates better prediction than other grains size curves. The median values are also in agreement with the statement since the medians are close to 1. This is further supported by having the predicted mean value closest to 1 for both models compared to other grains sizes considered.

Though the 100 µm curves show less uncertainty (relatively flat) in grain size estimation at shorter wavelengths by the ISM than the slab model, at longer wavelengths ($\mathrm{>}$3 µm) the scenario is opposite. Consequently, the 100 µm curves from the slab model show much better, overall, than the ISM's prediction for the corresponding water ice phases and temperatures. This is further supported by the median and means $\mathrm{\pm}$ 1$\sigma$ of predicted grain sizes for both ice phases by the approximation models at that grain size. The 1000 µm curve from the slab model for both ice phases show a scattered prediction of grains size among the model's predicted grain sizes -- also evident by having the highest median values among grain sizes by the slab model. The corresponding 1000 µm curve by ISM, in contrast, predicts much higher discrepancies.

For larger particles, the approximation models predict much smaller grain sizes at longer wavelengths (Fig. \ref{Fig2} and \ref{Fig3}). This is because the approximation models do not reproduce the characteristic higher saturation level of the Mie model \citep{hansen2009}. Accordingly, the difference in grain sizes for the 1000 µm curve by ISM is much higher than that of the 1 µm curve where Rayleigh effects render higher uncertainty in grain size prediction by the approximation model. The mean and median values of predicated grain sizes by the ISM's 1000 µm curve for both phases and every temperature are the largest among all ISM's predicted grain sizes.

\subsection{Chracteristic absorption coefficient}

\begin{figure*}[ht!]
\plotone{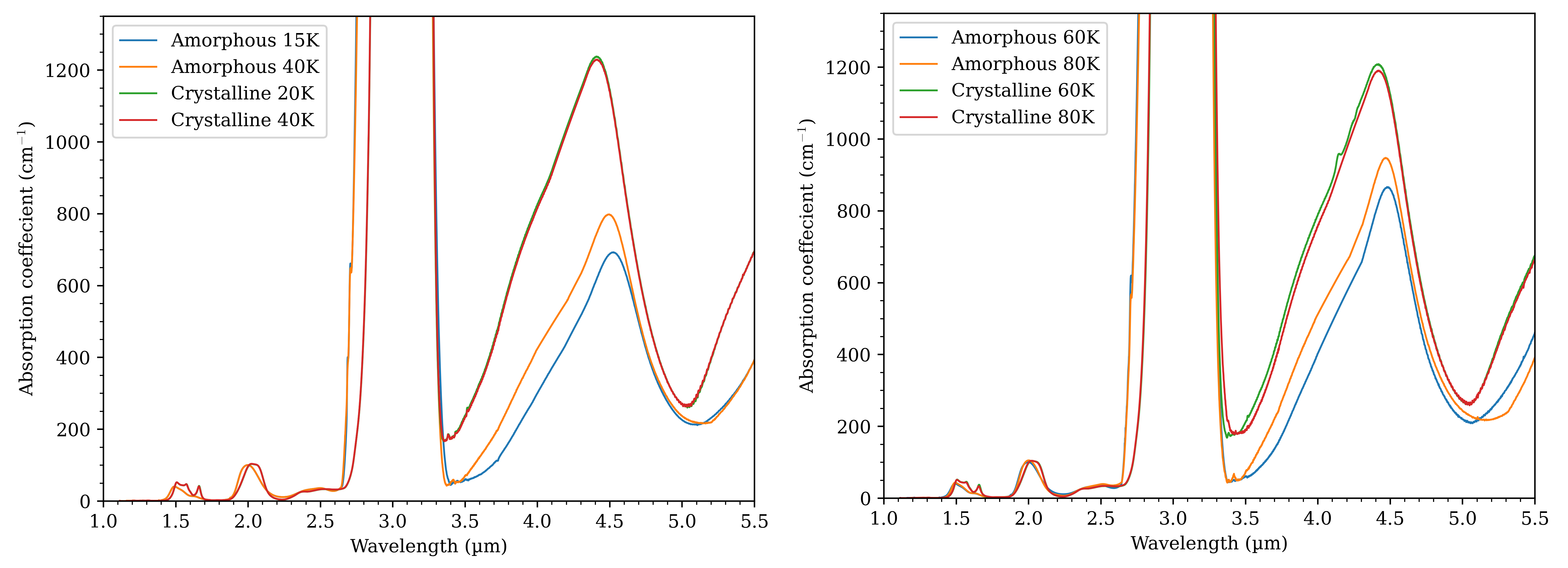}
\caption{Absorption coefficient of amorphous and crystalline H${}_{2}$O ice at different temperatures over NIR wavelengths. Both phases of H${}_{2}$O have a broad absorption coefficient peak of around 3.0 µm with a little shift toward shorter wavelengths by the Ia. Another absorption peak is seen at 4.5 µm for Ia while this peak shifts toward shorter wavelengths at 4.4 for Ic with much higher $\alpha$ values. The higher absorption peak for Ic compared to Ia at longer wavelengths is associated with the higher \textit{k }values at these wavelengths.
\label{Fig4}}
\end{figure*}

We plot the $\alpha$ distribution for Ia and Ic at different temperatures as a function of wavelengths (Fig. \ref{Fig4}). At every temperature, both phases have a broad and the highest $\alpha$ peak of around 3.0 µm wavelengths with a slight shift toward shorter wavelengths by the Ia than the Ic phase. The higher absorption peak at 3.0 µm is due to the higher \textit{k} values at these bands which makes the inability to gain size prediction by the approximation models. This evidence supports the assumption that the slab model should be used if \textit{k} $\mathrm{<}$ $\mathrm{<}$ 1 \citep{hapke1981}, and consequently a higher \textit{k} value leads to failure in predicting accurate grain size. A blue shift in $\alpha$ spectra of Ia compared to the Ic phase is also evident at the weak 2 µm band. This characteristic $\alpha$ distribution for H${}_{2}$O ice phases coincides with the fact that Ic has much stronger, sharper, and red-shifted at infrared bands than Ia \citep{schmitt1998}. 

Higher values of $\alpha$ are seen at the longer wavelengths for both ice phases. However, a peak of $\alpha$ happens around 4.5 µm for Ia while this peak shifts toward shorter wavelengths at 4.4 for Ic -- in agreement with \cite{mastrapa2009l}. The effects of these peaks are evident in the shapes of relative grain size curves for larger particles i.e., 100 and 1000 µm curves in Fig. \ref{Fig2} and \ref{Fig3}. Notably, the Ia shows comparatively much lower $\alpha$ values around the band of 4.5 µm region than the Ic at every investigated temperature. This relatively higher $\alpha$ value at this wavelength for Ic compared to Ia is also apparent in the relative grain-size curve of the 1000 µm particle (Fig. \ref{Fig2} and \ref{Fig3}) where the Ic reveals a higher inconsistency in grain size prediction than the Ia at this wavelength for both approximation models.

The $\alpha$ of pure CH${}_{4}$ and \textbf{N${}_{2}$:}CH${}_{4}$ (N${}_{2}$ saturated with CH${}_{4}$) systems, the other abundant volatile ices found on KBOs and TNOs surfaces, at NIR wavelengths (see Fig. 3 of \citealt{emran2022}) have a much lower $\alpha$ than water ice phases. A simple interpretation of this fact is that may be the H${}_{2}$O ice has higher absorption and lower reflectance compared to the pure CH${}_{4}$ and \textbf{N${}_{2}$:}CH${}_{4}$ ices on the TNOs (e.g., Triton) and KBOs (e.g., Pluto, Eris, and Sedna). The results above show that the higher discrepancies in predicated grains sizes for larger particles are largely aligned to the wavelengths with higher $\alpha$ values. Accordingly, the estimation of H${}_{2}$O ice grain size using the Hapke approximation models for larger particles may be susceptible to higher uncertainty than that of CH${}_{4}$ and \textbf{N${}_{2}$:}CH${}_{4}$ ices at the Kuiper belt \citep{emran2022}. 

The higher uncertainty in H${}_{2}$O ice than CH${}_{4}$ and \textbf{N${}_{2}$:}CH${}_{4}$ ices is further supported by the fact that the $\alpha$ of Ia at the 4.5 µm absorption band (and slightly at 1.5 and 2 µm bands) increases with the increase of temperature from 15 to 80 K (Fig. \ref{Fig4}). This $\alpha$ vs temperature characteristic of Ia may likely be the cause of the trend of higher discrepancies in grain size estimation with the higher temperatures at these bands (Fig. \ref{Fig2}). In contrast, Ic does not show any substantial change in $\alpha$ values at the 4.4 µm absorption band due to temperature changes, although $\alpha$ slightly decreases with temperature increase. Therefore, in the case of Ic, there are no substantial differences in grain size estimations at this band between the temperatures (Fig. \ref{Fig3}). However, we acknowledge that this statement on uncertainty in different grain size estimations of H${}_{2}$O, CH${}_{4}$, and \textbf{N${}_{2}$:}CH${}_{4}$ ices is purely based on the simplest interpretation from the comparison of the $\alpha$ values between the ice species at NIR bands, while other factors involved, the scenario may not be true.

\section{Discussion }

Water ice grain size estimation for outer solar system bodies from near-infrared wavelengths has reportedly been done using the optical constants of Ic and/or Ia by utilizing RTMs. However, estimation of grains sizes on planetary regolith depends primarily on the calculation of \textit{w} rather than a choice of bidirectional scattering models \citep{hansen2009}. The Hapke approximation models are a well-recognized method for deriving the \textit{w} and, therefore, the models have been used in estimating surface grain sizes of icy bodies. We assess the relative grain size prediction by widely used Hapke approximation models for water ices phases found in the outer solar system and beyond. Accordingly, we use a range of fixed grains sizes (1, 10, 100, and 1000 µm) in Mie calculation and then determine the corresponding grain sizes from the Hapke approximation models to fit the Mie spectra at different NIR wavelengths.

We use the Hapke models for predicting grain sizes of a 1 µm radii even though the approximation model assumes that the particle size should be much larger than the wavelengths \citep{hapke1981}. This study also uses wavelengths close to or even larger than a grain size of 1 µm. Indeed, we test a wide range of grain sizes where the 1 µm grain size is just an instance, while particles with sizes much larger than the wavelengths were also considered. The rationale for using 1 µm radii is that there has been reported the presence of H${}_{2}$O ice with similar sizes on the surface of outer solar system bodies such as KBOs \citep[e.g.,][]{cook2019}. Moreover, the use of grain size close to 1 µm emphasizes the Rayleigh effect that happens at grains that have sizes close to wavelengths for the outer solar system surface ices \citep[e.g.,][]{hansen2009, emran2022}. And by taking smaller particle sizes, we evaluate the grain size prediction discrepancies at wavelengths close to or even larger than the grain size of 1 µm. On top of that our result shows how deviant, as it is expected, grain size prediction the approximation models can produce (Fig. \ref{Fig2} and \ref{Fig3}) when the wavelengths are close to or even larger than the grain size. 

The predicted grain sizes by the Hapke approximation models vary over the investigated wavelengths due to the optical constants (and, therefore, different absorption coefficient values) at different NIR bands and the formulation of the approximation models used (see Eq. \ref{Eq.2} and \ref{Eq.5}). In simple words, the approximation models predict different grain sizes based on optical constant values (\textit{n} + \textit{ik}) at different wavelengths. Accordingly, assessing the predictions of these varying grain sizes at different wavelengths by the approximation models was the motivation of the study and thus, we evaluate the discrepancies of predicting varying grain sizes by the scattering models. As mentioned above, we only consider the optical constant of cubic crystalline ice as a generalization of crystalline ice. However, there are only subtle differences in spectra between the cubic and hexagonal phases \citep{bertie1967}, and thus, we emphasize that the crystalline phases are substitutable \citep{mastrapa2009l} and the use of Ic for generalization of the crystalline ice is reasonable.

Between the ice phases, the grain size estimation for Ic using the Hapke approximation models results in a much closer, in general, estimate to Mie's prediction. Conversely, the Hapke approximation models show higher discrepancies in grain size estimation for Ia. The selection of temperatures for the ice phases is also a consideration. The discrepancy in grain size estimation of larger Ia particles increases with temperatures from 15 to 80 K at longer wavelengths, whereas these discrepancies in grain size estimation are negligible for Ic between the temperature range. However, grain size estimation for larger particles at the broad 4.4 - 4.5 µm absorption band leads to higher discrepancy for Ic compared to the Ia. The relative discrepancies of the grain size estimation is better (close to 1) for the Hapke slab model than ISM at cryogenic temperatures. This indicates Hapke slab is a better approximation model for deriving the water ice grain sizes in the icy bodies in the far reach of the solar system. Our result is consistent with the results of previously analyzed \cite{hansen2009} using the optical constant of Ic at a temperature of 110 K.

Owing to the Rayleigh effects on single scattering albedo, neither of the approximation models predicts accurate grain size estimation results for the particles with diameters less than or close to the wavelengths ($\mathrm{\le}$ 5 µm). For both Ia and Ic, grain size estimations for 10 µm radii H${}_{2}$O ice are best predicted by both Hapke approximation models. Though grain size prediction by the approximation models for larger particles shows higher discrepancies largely at the wavelengths with higher absorption coefficient values, the uncertainty in the instance of ISM is much higher than the Hapke slab model. The higher discrepancies of grain size estimation for larger particles by ISM can be compromised by adjusting the free parameters used in calculating single scattering albedo, for instance, the study of \cite{roush2007} used a higher value of \textit{s} to fit modeled spectra. 

While formulating the ``\textit{espat}'' function for the slab model, \cite{hapke1981} used a scaling factor of \textit{\^{g}} = 2/3 for particle diameter. However, the scaling factor of diameter in the Hapke slab model varies between 3/4 - 4/3 which can, in principle, result in a difference in grain size estimation by the slab model \citep{hansen2009}. Likewise, the difference in grain size estimation by ISM can also be produced by scaling factor to the particle diameter through calculating the mean free path of the photon $\mathrm{\langle }$\textit{D}$\mathrm{\rangle }$ as seen in Eq. \ref{Eq.7}. There have been different approximate expressions for the relationship between \textit{D} and $\mathrm{\langle }$\textit{D}$\mathrm{\rangle }$. This relationship is approximated as \textit{$\langle $D$\rangle $} $\mathrm{\cong }$ 0.9\textit{D} or 2\textit{D}/3 \citep[e.g.,][]{hapke2012, melamed1963} for spherical particles while it's \textit{$\langle $D$\rangle $ }= 0.2D for irregular particles \citep{shkuratov2005}. Hence, the grain sizes estimation by the approximations models can be improved by carefully adjusting the scale factor for particle diameters. We use a scale factor of 1 for the slab model while the mean free path of a photon in ISM was calculated as a function of the diameter and real part of the refraction index.

An obvious difference between the ISM and the slab models is the calculation of average Fresnel reflection coefficients -- the former model employs data from both \textit{n} and \textit{k} parts while the latter uses only \textit{n }part of the refractive indices \citep[e.g.,][]{emran2022}. Since the slab model can be applied when \textit{k} is very small \citep{hapke1981, hapke1993} and therefore, at wavelengths with a substantially higher \textit{k} value can lead to a higher difference in calculated results between the models. Note that if the approximation model assumes the internal scattering coefficient, \textit{s} = 0 (as used by many existing studies, for example, \citealt{lucey1998, lawrence2007}) then the internal diffused reflectance, \textit{r }equals to be\textit{ }zero as well. However, \cite{sharkey2019} expressed the relationships between the \textit{s} and \textit{D} (``effective diameter'') as \textit{s} = 1/\textit{D }and the number of scattering events (\textit{sD}) = 1 within a single grain\textit{. }Thus, this relationship indicates that \textit{s }cannot be zero \citep{emran2022}.

As has previously been mentioned that Mie's theory \citep{mie1908} can accurately calculate the scattering properties of uniform (shapes) particles. Further, besides application to scattering properties for the equant particles of Mie spheres, the Mie scattering model has been devised to be satisfactorily employed for particles with varied ``equivalent'' non-spherical shapes if each non-spherical particle can be represented by assembles of spheres with the same volume-to-surface-area ratio as the non-spherical particle \citep[e.g.,][]{grenfell1999, grenfell2005}. In other words, roughly similar scattering properties have been reported by irregular particles to that of the scattering properties calculated by the similar-sized spherical particles using Mie's theory \citep{hapke1981}. This implies that the scattering properties of Mie theory for spherical particles can be mimicked to irregular particles \citep{neshyba2003}, and thus application to Mie theory in the grain size estimation of a Mie's equant particle is somewhat analogous to non-spherical/ irregular grains \citep[e.g.,][]{hapke1981, hansen2009}. 

While both Hapke approximation models do not account for the Rayleigh effect on the single scattering albedo that happens to grain size close to the wavelengths ($\mathrm{\le}$5 µm), the Mie theory appropriately accounts for this shortcoming. Though the rough average difference between the predicted grain sizes by the ISM and slab model is about $\mathrm{\sim}$ 10 -- 20\% (based on the median values in Table. \ref{tab:1}) for a wide range of particles, we emphasize that it's still an over/underestimation by one approximation model to another. And our study shows which approximation model predicts much better results than others in estimating the grains sizes of Ia and Ic found in the vast region of the outer solar system. A better estimation of H${}_{2}$O ice grain sizes by the slab model than ISM is also in agreement with the grain size estimation of the abundant surface volatile of N${}_{2}$ and CH${}_{4}$ ices on TNOs and KBOs surfaces \citep{emran2022}. Thus, the Hapke slab model is better equipped for the implementation of RTMs in the approximate grain size estimation for a range of surface ices (H${}_{2}$O, N${}_{2}$, and CH${}_{4}$) in the outer solar system bodies.

We recommend using the Mie formulation of the single scattering albedo for estimation of H${}_{2}$O ice (Ia and Ic) grain sizes for unknown spectra of outer solar system bodies and beyond. When choosing the approximation models, we recommend using the Hapke slab model over ISM since our result concludes that the slab model better approximates the H${}_{2}$O grain size prediction, overall, than the internal scattering model at 1 -- 5 µm wavelengths. Note that the single scattering albedo may not be readily available from the measurements of spacecraft observations. However, the radiance or reflectance measurement by spacecraft observation can be converted to single scattering albedo using the radiative transfer model of \cite{hapke1993}.

\section{Conclusion}
The mechanical strengths and thermal properties (e.g., temperatures) of planetary surfaces are directly influenced by the grain size of planetary regolith \citep[e.g.,][]{gundlach2013}. Thus, an accurate grain size estimation of the outer solar system bodies is very important for an improved understanding of the surface volatile transport and thermophysical modeling of the planetary bodies. The icy bodies in the outer solar system host a widespread presence of water ice on their surfaces. Understanding the grain sizes of planetary regolith supports landing or sample return missions \citep{gundlach2013}, and thus accurate information about the water ice grains sizes is crucial for a future mission to the outer reach of the solar system. 

We assess the discrepancies in the grain size estimation of amorphous and crystalline H${}_{2}$O ice at cryogenic temperatures from 15 -- 80 K at a $\mathrm{\sim}$20 K interval. The relative grain size prediction by the Hapke approximation models was assessed concerning ice phases, grain sizes, and temperatures relevant to the icy zones of the solar system and the interstellar space. Having a satisfactory application of Mie theory to an equant as well as ``equivalent'' non-spherical particles, we recommend using Mie formulation for unknown spectra of the outer solar system bodies. However, if approximation models are opted to use, our study favors using the slab model as a better option than the internal scattering model.

We demonstrate which model works better for the approximate grain size estimation of phases of water ice found in the outer solar system. Incorporation with the results from \cite{emran2022}, we conclude which approximate model is better in predicting the grain sizes of a variety of surface ices found in the outer solar system bodies, more specifically on trans-Neptunian objects and Kuiper belt objects. Thus, this study put forward a prescription for future studies in choosing scattering models in estimating the water ice grain sizes in the outer solar system and beyond using the radiative transfer models at near-infrared bands.


\begin{acknowledgments}
The authors would like to thank anonymous reviewers for their useful comments.
\end{acknowledgments}

\software{NumPy \citep{harris2020array}, SciPy \citep{virtanen2020scipy}, MatplotLib \citep{hunter2007matplotlib}, MiePython \break
\url{https://pypi.org/project/miepython/}}

\begin{appendix}

\section{Statistics}
The mean  and $\mathrm{\pm}$ 1\textit{$\sigma$} standard deviation of relative discrepancies in grain sizes are predicated by the approximation models

\begin{table*}[!htbp]
\begin{center}
\centering

\large

\caption{The statistics - mean  and $\mathrm{\pm}$ 1\textit{$\sigma$} standard deviation - of relative discrepancies in predicted grain sizes using the Hapke slab and ISM calculated at particle radii of 1, 10, 100, and 1000 µm for amorphous (a) and crystalline (b) H${}_{2}$O ice at different temperatures. \label{tab:2}}

\begin{tabular}{lcccccccc}


\multicolumn{9}{l}{a.}\\
\hline

\multirow{3}{*}{\rotatebox{90}{
\makecell{Grain radii \hspace{.05in} \\ (µm)}}} & \multicolumn{8}{c}{\textbf{Amorphous H${}_{2}$O ice${}^{*}$}}  \\ [6pt]
\cline{2-9} 
 & \multicolumn{2}{c}{15 K} & \multicolumn{2}{c}{40 K} & \multicolumn{2}{c}{60 K} & \multicolumn{2}{c}{80 K} \\ [6pt]
\cline{2-9}
 & Slab & ISM & Slab & ISM & Slab & ISM & Slab & ISM \\  [6pt] \hline  
1 & 1.89$\mathrm{\pm}$3.06 & 2.07$\mathrm{\pm}$2.96 & 1.96$\mathrm{\pm}$3.42 & 2.25$\mathrm{\pm}$4.13 & 2.08$\mathrm{\pm}$3.97 & 2.43$\mathrm{\pm}$5.14\newline  & 2.11$\mathrm{\pm}$4.10 & 2.44$\mathrm{\pm}$5.22 \\ \hline 
10 & 1.21$\mathrm{\pm}$0.74 & 1.40$\mathrm{\pm}$0.92 & 1.23$\mathrm{\pm}$0.82 & 1.14$\mathrm{\pm}$0.98 & 1.26$\mathrm{\pm}$0.95 & 1.42$\mathrm{\pm}$1.06 & 1.27$\mathrm{\pm}$0.98 & 1.43$\mathrm{\pm}$1.10 \\ \hline 
100 & 1.58$\mathrm{\pm}$0.72 & 2.05$\mathrm{\pm}$1.62 & 1.57$\mathrm{\pm}$0.75 & 2.06$\mathrm{\pm}$1.59 & 1.59$\mathrm{\pm}$0.80 & 2.04$\mathrm{\pm}$1.57 & 1.59$\mathrm{\pm}$0.81 & 2.06$\mathrm{\pm}$1.59 \\ \hline 
1000 & 1.51$\mathrm{\pm}$1.00 & 2.62$\mathrm{\pm}$2.44 & 1.51$\mathrm{\pm}$1.02 & 2.57$\mathrm{\pm}$2.47 & 1.52$\mathrm{\pm}$1.03 & 2.58$\mathrm{\pm}$2.48 & 1.50$\mathrm{\pm}$0.99 & 2.41$\mathrm{\pm}$2.29  \\ \hline 
& & & & & & & & \\

\multicolumn{9}{l}{ b. }\\
\cline{1-9}

\multirow{3}{*}{\rotatebox{90}{
\makecell{Grain radii \hspace{.05in} \\ (µm)}}} & \multicolumn{8}{c}{\textbf{Crystalline H${}_{2}$O ice${}^{*}$}}  \\ [6pt]
\cline{2-9} 
 & \multicolumn{2}{c}{20 K} & \multicolumn{2}{c}{40 K} & \multicolumn{2}{c}{60 K} & \multicolumn{2}{c}{80 K} \\ [6pt]
\cline{2-9}
 & Slab & ISM & Slab & ISM & Slab & ISM & Slab & ISM \\  [6pt] \hline 
1 & 2.07$\mathrm{\pm}$5.76 & 2.19$\mathrm{\pm}$5.99 & 2.05$\mathrm{\pm}$5.64 & 2.17$\mathrm{\pm}$5.89 & 1.97$\mathrm{\pm}$5.15 & 2.13$\mathrm{\pm}$5.62 & 1.97$\mathrm{\pm}$5.15 & 2.13$\mathrm{\pm}$5.62 \\ \hline 
10 & 1.24$\mathrm{\pm}$1.35 & 1.32$\mathrm{\pm}$1.08 & 1.24$\mathrm{\pm}$1.3 & 1.31$\mathrm{\pm}$1.05 & 1.22$\mathrm{\pm}$1.20 & 1.30$\mathrm{\pm}$0.99 & 1.22$\mathrm{\pm}$1.20 & 1.30$\mathrm{\pm}$0.99 \\ \hline 
100 & 1.44$\mathrm{\pm}$1.05 & 1.74$\mathrm{\pm}$1.25 & 1.44$\mathrm{\pm}$1.04 & 1.74$\mathrm{\pm}$1.24 & 1.43$\mathrm{\pm}$0.99 & 1.75$\mathrm{\pm}$1.26 & 1.43$\mathrm{\pm}$0.99 & 1.75$\mathrm{\pm}$1.26 \\ \hline 
1000 & 1.28$\mathrm{\pm}$0.86 & 2.30$\mathrm{\pm}$2.21 & 1.27$\mathrm{\pm}$0.85 & 2.31$\mathrm{\pm}$2.21 & 1.27$\mathrm{\pm}$0.84 & 2.31$\mathrm{\pm}$2.22 & 1.22$\mathrm{\pm}$0.84 & 2.31$\mathrm{\pm}$2.22 \\ \hline

\end{tabular}

\end{center}

\tablecomments{The statistics were computed by normalizing the estimated grain sizes from the Hapke slab and ISM to the Mie grain size. The number values are the mean  and $\mathrm{\pm}$ 1\textit{$\sigma$} standard deviation -- representing how many times of predicted grain sizes by slab and ISM models to the Mie grain size H${}_{2}$O ice phases at different temperatures}

\end{table*}

\section{Notation}
\label{Notation}

List of notation and symbols used in this paper

\textit{$\langle $D$\rangle$}  mean free path of a photon 

\textit{D} particle diameter

\textit{D${}_{e}$} effective particle sizes

\textit{$\xi$} asymmetry parameters of Mie theory

\textit{\^{g}} a constant, roughly = 1

\textit{k} imaginary part of the refractive index

\textit{n} real part of the refractive index

\textit{r }internal diffused reflectance

\textit{s }internal scattering coefficient

\textit{S${}_{e}$ }Fresnel reflection coefficient for externally incident light,\textit{}

\textit{S${}_{i}$ }Fresnel reflection coefficient for internally incident light,

\textit{w} single scattering albedo

\textit{w' $\delta$}-Eddington Mie single scattering albedo\textit{ }

\textit{$\alpha$} absorption coefficient

\textit{$\lambda$} wavelength

\textrm{$\Theta$} internal transmission coefficient

\end{appendix}


\bibliography{sample631}{}

\begin{thebibliography}{}
\expandafter\ifx\csname natexlab\endcsname\relax\def\natexlab#1{#1}\fi
\providecommand{\url}[1]{\href{#1}{#1}}
\providecommand{\dodoi}[1]{doi:~\href{http://doi.org/#1}{\nolinkurl{#1}}}
\providecommand{\doeprint}[1]{\href{http://ascl.net/#1}{\nolinkurl{http://ascl.net/#1}}}
\providecommand{\doarXiv}[1]{\href{https://arxiv.org/abs/#1}{\nolinkurl{https://arxiv.org/abs/#1}}}

\bibitem[{Baragiola(2003)}]{baragiola2003}
Baragiola, R.~A. 2003, Planetary and Space Science, 51, 953

\bibitem[{Barucci \& Merlin(2020)}]{barucci2020}
Barucci, M.~A., \& Merlin, F. 2020, in The Trans-Neptunian Solar System
  (Elsevier), 109--126

\bibitem[{Bertie \& Whalley(1967)}]{bertie1967}
Bertie, J., \& Whalley, E. 1967, The Journal of chemical physics, 46, 1271

\bibitem[{Chen \& Stadtherr(1981)}]{chen1981}
Chen, H.-S., \& Stadtherr, M.~A. 1981, Computers \& Chemical Engineering, 5,
  143

\bibitem[{Cook {et~al.}(2019)Cook, Dalle~Ore, Protopapa, Binzel, Cruikshank,
  Earle, Grundy, Ennico, Howett, Jennings, {et~al.}}]{cook2019}
Cook, J.~C., Dalle~Ore, C.~M., Protopapa, S., {et~al.} 2019, Icarus, 331, 148

\bibitem[{Cruikshank {et~al.}(1998)Cruikshank, Roush, Owen, Quirico, \&
  Bergh}]{cruikshank1998}
Cruikshank, D.~P., Roush, T.~L., Owen, T.~C., Quirico, E., \& Bergh, C.~D.
  1998, in Solar system ices (Springer), 655--684

\bibitem[{Cruikshank {et~al.}(2000)Cruikshank, Schmitt, Roush, Owen, Quirico,
  Geballe, de~Bergh, Bartholomew, Dalle~Ore, Dout{\'e},
  {et~al.}}]{cruikshank2000}
Cruikshank, D.~P., Schmitt, B., Roush, T.~L., {et~al.} 2000, Icarus, 147, 309

\bibitem[{Dalle~Ore {et~al.}(2018)Dalle~Ore, Protopapa, Cook, Grundy,
  Cruikshank, Verbiscer, Ennico, Olkin, Stern, Weaver, {et~al.}}]{dalle2018}
Dalle~Ore, C.~M., Protopapa, S., Cook, J., {et~al.} 2018, Icarus, 300, 21

\bibitem[{Dumas {et~al.}(2007)Dumas, Merlin, Barucci, De~Bergh, Hainault,
  Guilbert, Vernazza, \& Doressoundiram}]{dumas2007}
Dumas, C., Merlin, F., Barucci, M., {et~al.} 2007, Astronomy \& Astrophysics,
  471, 331

\bibitem[{Earle {et~al.}(2017)Earle, Binzel, Young, Stern, Ennico, Grundy,
  Olkin, Weaver, {et~al.}}]{earle2017}
Earle, A.~M., Binzel, R.~P., Young, L.~A., {et~al.} 2017, Icarus, 287, 37

\bibitem[{Emran \& Chevrier(2022)}]{emran2022}
Emran, A., \& Chevrier, V. 2022, arXiv preprint arXiv:2110.14591

\bibitem[{Ferriere(2001)}]{ferriere2001}
Ferriere, K.~M. 2001, Reviews of Modern Physics, 73, 1031

\bibitem[{Filacchione {et~al.}(2012)Filacchione, Capaccioni, Ciarniello, Clark,
  Cuzzi, Nicholson, Cruikshank, Hedman, Buratti, Lunine,
  {et~al.}}]{filacchione2012}
Filacchione, G., Capaccioni, F., Ciarniello, M., {et~al.} 2012, Icarus, 220,
  1064

\bibitem[{Grenfell {et~al.}(2005)Grenfell, Neshyba, \& Warren}]{grenfell2005}
Grenfell, T.~C., Neshyba, S.~P., \& Warren, S.~G. 2005, Journal of Geophysical
  Research: Atmospheres, 110

\bibitem[{Grenfell \& Warren(1999)}]{grenfell1999}
Grenfell, T.~C., \& Warren, S.~G. 1999, Journal of Geophysical Research:
  Atmospheres, 104, 31697

\bibitem[{Grundy \& Schmitt(1998)}]{grundy&schmitt1998}
Grundy, W., \& Schmitt, B. 1998, Journal of Geophysical Research: Planets, 103,
  25809

\bibitem[{Grundy \& Fink(1991)}]{grundy1991}
Grundy, W.~M., \& Fink, U. 1991, Icarus, 93, 379

\bibitem[{Gundlach \& Blum(2013)}]{gundlach2013}
Gundlach, B., \& Blum, J. 2013, Icarus, 223, 479

\bibitem[{Hansen(2009)}]{hansen2009}
Hansen, G. 2009, Icarus, 203, 672

\bibitem[{Hansen \& McCord(2004)}]{hansen2004}
Hansen, G.~B., \& McCord, T.~B. 2004, Journal of Geophysical Research: Planets,
  109

\bibitem[{Hapke(1981)}]{hapke1981}
Hapke, B. 1981, Journal of Geophysical Research: Solid Earth, 86, 3039

\bibitem[{Hapke(1993)}]{hapke1993}
---. 1993, Theory of reflectance and emittance spectroscopy: Topics in Remote
  Sensing (Cambridge university press)

\bibitem[{Hapke(2012)}]{hapke2012}
---. 2012, Theory of reflectance and emittance spectroscopy (Cambridge
  university press)

\bibitem[{Harris {et~al.}(2020)Harris, Millman, Van Der~Walt, Gommers,
  Virtanen, Cournapeau, Wieser, Taylor, Berg, Smith,
  {et~al.}}]{harris2020array}
Harris, C.~R., Millman, K.~J., Van Der~Walt, S.~J., {et~al.} 2020, Nature, 585,
  357

\bibitem[{Herbst(2001)}]{herbst2001}
Herbst, E. 2001, Chemical Society Reviews, 30, 168

\bibitem[{Hudgins {et~al.}(1993)Hudgins, Sandford, Allamandola, \&
  Tielens}]{hudgins1993}
Hudgins, D., Sandford, S., Allamandola, L., \& Tielens, A. 1993, The
  Astrophysical Journal Supplement Series, 86, 713

\bibitem[{Hunter(2007)}]{hunter2007matplotlib}
Hunter, J.~D. 2007, Computing in science \& engineering, 9, 90

\bibitem[{Joseph {et~al.}(1976)Joseph, Wiscombe, \& Weinman}]{joseph1976}
Joseph, J.~H., Wiscombe, W., \& Weinman, J. 1976, Journal of Atmospheric
  Sciences, 33, 2452

\bibitem[{Kofman {et~al.}(2019)Kofman, He, ten Kate, \& Linnartz}]{kofman2019}
Kofman, V., He, J., ten Kate, I.~L., \& Linnartz, H. 2019, The Astrophysical
  Journal, 875, 131

\bibitem[{Lawrence \& Lucey(2007)}]{lawrence2007}
Lawrence, S.~J., \& Lucey, P.~G. 2007, Journal of Geophysical Research:
  Planets, 112

\bibitem[{Li \& Li(2011)}]{li2011}
Li, S., \& Li, L. 2011, Journal of Geophysical Research: Planets, 116

\bibitem[{Lucey(1998)}]{lucey1998}
Lucey, P.~G. 1998, Journal of Geophysical Research: Planets, 103, 1703

\bibitem[{Mastrapa {et~al.}(2008)Mastrapa, Bernstein, Sandford, Roush,
  Cruikshank, \& Dalle~Ore}]{mastrapa2008}
Mastrapa, R., Bernstein, M., Sandford, S., {et~al.} 2008, Icarus, 197, 307

\bibitem[{Mastrapa {et~al.}(2009)Mastrapa, Sandford, Roush, Cruikshank, \&
  Dalle~Ore}]{mastrapa2009l}
Mastrapa, R., Sandford, S., Roush, T., Cruikshank, D., \& Dalle~Ore, C. 2009,
  The Astrophysical Journal, 701, 1347

\bibitem[{Melamed(1963)}]{melamed1963}
Melamed, N. 1963, Journal of Applied Physics, 34, 560

\bibitem[{Merlin {et~al.}(2010)Merlin, Barucci, de~Bergh, Fornasier,
  Doressoundiram, Perna, \& Protopapa}]{merlin2010}
Merlin, F., Barucci, M., de~Bergh, C., {et~al.} 2010, Icarus, 208, 945

\bibitem[{Mie(1908)}]{mie1908}
Mie, G. 1908, Annalen der physik, 330, 377

\bibitem[{Mustard \& Glotch(2019)}]{mustard&glotach2019}
Mustard, J.~F., \& Glotch, T.~D. 2019, Remote Compositional Analysis:
  Techniques for Understanding Spectroscopy, Mineralogy, and Geochemistry of
  Planetary Surfaces, 21

\bibitem[{Neshyba {et~al.}(2003)Neshyba, Grenfell, \& Warren}]{neshyba2003}
Neshyba, S.~P., Grenfell, T.~C., \& Warren, S.~G. 2003, Journal of Geophysical
  Research: Atmospheres, 108

\bibitem[{Pollack {et~al.}(1994)Pollack, Hollenbach, Beckwith, Simonelli,
  Roush, \& Fong}]{pollack1994}
Pollack, J.~B., Hollenbach, D., Beckwith, S., {et~al.} 1994, The Astrophysical
  Journal, 421, 615

\bibitem[{Powell(1970)}]{powell1970}
Powell, M.~J. 1970, Numerical methods for nonlinear algebraic equations

\bibitem[{Protopapa {et~al.}(2017)Protopapa, Grundy, Reuter, Hamilton,
  Dalle~Ore, Cook, Cruikshank, Schmitt, Philippe, Quirico,
  {et~al.}}]{protopapa2017}
Protopapa, S., Grundy, W., Reuter, D., {et~al.} 2017, Icarus, 287, 218

\bibitem[{Raponi {et~al.}(2016)Raponi, Ciarniello, Capaccioni, Filacchione,
  Tosi, De~Sanctis, Capria, Barucci, Longobardo, Palomba,
  {et~al.}}]{raponi2016}
Raponi, A., Ciarniello, M., Capaccioni, F., {et~al.} 2016, Monthly Notices of
  the Royal Astronomical Society, 462, S476

\bibitem[{Roush(1994)}]{roush1994}
Roush, T.~L. 1994, Icarus, 108, 243

\bibitem[{Roush {et~al.}(2007)Roush, Esposito, Rossman, \&
  Colangeli}]{roush2007}
Roush, T.~L., Esposito, F., Rossman, G.~R., \& Colangeli, L. 2007, Journal of
  Geophysical Research: Planets, 112

\bibitem[{Savitzky \& Golay(1964)}]{savitzky1964}
Savitzky, A., \& Golay, M.~J. 1964, Analytical chemistry, 36, 1627

\bibitem[{Schmitt {et~al.}(1998)Schmitt, Quirico, Trotta, \&
  Grundy}]{schmitt1998}
Schmitt, B., Quirico, E., Trotta, F., \& Grundy, W. 1998, in Solar System Ices
  (Springer), 199--240

\bibitem[{Sharkey {et~al.}(2019)Sharkey, Reddy, Sanchez, Izawa, \&
  Emery}]{sharkey2019}
Sharkey, B.~N., Reddy, V., Sanchez, J.~A., Izawa, M.~R., \& Emery, J.~P. 2019,
  The Astronomical Journal, 158, 204

\bibitem[{Shepard(2017)}]{shepard2017}
Shepard, M.~K. 2017, Introduction to planetary photometry (Cambridge University
  Press)

\bibitem[{Shepard \& Helfenstein(2007)}]{shepard2007}
Shepard, M.~K., \& Helfenstein, P. 2007, Journal of Geophysical Research:
  Planets, 112

\bibitem[{Shkuratov \& Grynko(2005)}]{shkuratov2005}
Shkuratov, Y.~G., \& Grynko, Y.~S. 2005, Icarus, 173, 16

\bibitem[{Tegler {et~al.}(2010)Tegler, Cornelison, Grundy, Romanishin,
  Abernathy, Bovyn, Burt, Evans, Maleszewski, Thompson, {et~al.}}]{tegler2010}
Tegler, S.~C., Cornelison, D., Grundy, W., {et~al.} 2010, The Astrophysical
  Journal, 725, 1296

\bibitem[{Virtanen {et~al.}(2020)Virtanen, Gommers, Oliphant, Haberland, Reddy,
  Cournapeau, Burovski, Peterson, Weckesser, Bright,
  {et~al.}}]{virtanen2020scipy}
Virtanen, P., Gommers, R., Oliphant, T.~E., {et~al.} 2020, Nature methods, 17,
  261

\bibitem[{Wiscombe(1979)}]{wiscombe1979}
Wiscombe, W.~J. 1979, Mie scattering calculations: Advances in technique and
  fast, vector-speed computer codes, Vol.~10 (National Technical Information
  Service, US Department of Commerce)

\bibitem[{Wiscombe(1980)}]{wiscombe1980}
---. 1980, Applied optics, 19, 1505

\bibitem[{Wiscombe \& Warren(1980)}]{wiscombe&warren1980}
Wiscombe, W.~J., \& Warren, S.~G. 1980, Journal of Atmospheric Sciences, 37,
  2712

\end{thebibliography}
\bibliographystyle{aasjournal}



\end{document}